\documentclass[12pt]{spieman}  
\usepackage{amsmath,amsfonts,amssymb}
\usepackage{graphicx}
\usepackage{setspace}
\usepackage{tocloft}
\usepackage{tabularx}
\usepackage{hyperref}
\usepackage{mathrsfs}
\usepackage{booktabs}

\title{NeOTF: Guidestar-free neural representation for broadband dynamic imaging through scattering}

\author[1,*]{Yunong Sun}
\author[1,2,*]{Fei Xia}

\affil[1]{Department of Electrical Engineering and Computer Science, University of California, Irvine, CA 92697, USA}
\affil[2]{Beckman Laser Institute, University of California, Irvine, CA 92697, USA}

\cftpagenumbersoff{figure}
\cftpagenumbersoff{table} 
\begin{document} 
\maketitle

 \begin{abstract}
Dynamic imaging through time-varying scattering media is ubiquitous in real-world settings, yet it remains a defining unsolved problem as rapid spatiotemporal fluctuations overwhelm standard reconstruction pipelines that often rely on speckles with high signal-to-noise ratio. Existing approaches fall into two categories. Guidestar-based methods employ a guidestar to recover the system transfer function; however, in dynamic media, the speckle decorrelates rapidly, making the calibration quickly invalid. Guidestar-free methods infer information from speckle statistics, but rapid changes and noise often break phase retrieval. To overcome these limitations, we introduce NeOTF, a guidestar-free and neural-representation-based OTF retrieval method that enables dynamic imaging through time-varying scattering media. By optimizing this neural representation with only a few speckle images from unknown objects, NeOTF robustly retrieves the system's OTF without a guidestar. We experimentally demonstrate robust dynamic imaging through scattering with NeOTF at extremely low signal-to-noise ratio and broadband incoherent illumination (up to 300~nm spectral bandwidth) scenarios, and we numerically validate its dynamic imaging performance in time-varying scattering media leveraging spatio-temporal memory effect. Finally, we discuss and analyze its computational efficiency and generalization capabilities across anisotropic scattering media. These results establish NeOTF's promise as a practical and robust solution for dynamic imaging through scattering media. Open-sourced code and models are available at \href{https://github.com/Xia-Research-Lab/NeOTF}{https://github.com/Xia-Research-Lab/NeOTF}.
\end{abstract}


{\noindent \footnotesize\textbf{*}Yunong Sun,  \linkable{yunongs1@uci.edu} }
{\noindent \footnotesize\textbf{*}Fei Xia,  \linkable{fei.xia@uci.edu} }

\begin{spacing}{2}   

\section{Introduction}
\label{sect:intro}  
\begin{figure}[!htb]
    \centering
    \includegraphics[width=1.0\linewidth]{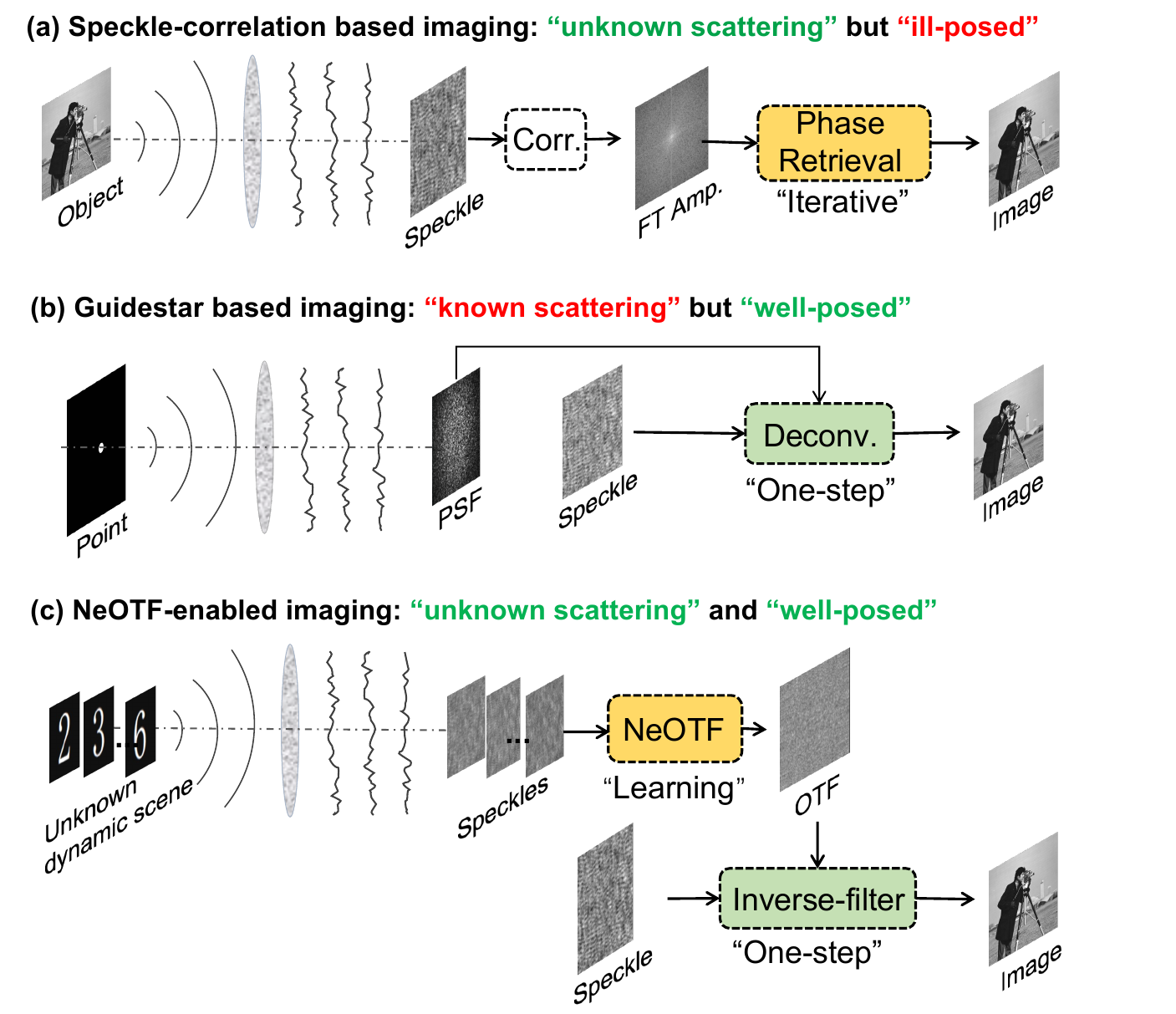}
    \caption{Illustration of main methods for imaging through scattering media. (a) Traditional single-shot speckle correlation method for imaging through a scattering medium. (b) Guidestar-based imaging method. The objects are directly recovered by deconvolution with the PSF calibrated from the point source which acts as a guidestar. (c) Our proposed image-guided NeOTF-based imaging method. The phase of the OTF is retrieved from the unknown speckles using iterative optimization. The image can be directly restored from the speckle by inverse filtering.}
    \label{fig:principle-illustrates}
\end{figure}

Imaging through scattering media is a fundamental challenge because scattering distorts and obscures the information carried by light, leading to severe degradation of image quality. Overcoming this challenge is crucial for a wide range of applications, including deep-tissue biomedical imaging, long-range remote sensing with lidar, and astronomical observation \cite{faccio2020non, yoon2020deep,bertolotti2022imaging}. To address this long-standing challenge, various techniques have been developed, which can be broadly categorized into several families. On the hardware side, it includes wavefront shaping methods that actively compensate spatially for aberrations and scattering \cite{popoff2010measuring, katz2012looking, yeminy2021guidestar, feng2023neuws, mididoddi2025threading}, as well as temporal gating methods that isolate ballistic photons by photon arrival time\cite{velten2012recovering, wu2021non}. On the computational side, it involves speckle correlation methods that exploit speckles' intensity distribution \cite{bertolotti2012non, li2018deep}. Among these, approaches leveraging the speckle correlation have gathered considerable attention\cite{bertolotti2012non, katz2014non}, as they enable computational imaging with speckle patterns without requiring complex hardware or direct access to the scattering medium (see Fig.~\ref{fig:principle-illustrates}(a)). Such methods assume the speckle images are acquired within the so-called memory-effect range~\cite{freund1988memory}, where the speckle patterns remain correlated. 

Speckle correlation techniques operate by exploiting the correlation properties of speckle patterns to retrieve the object's 2D spatial autocorrelation pattern. From this autocorrelation, the final image is typically reconstructed using iterative phase retrieval algorithms \cite{fienup1982phase}. However, the performance of these classical algorithms often hits a bottleneck: it is by nature an ill-posed problem that suffers from slow convergence and tendency to become trapped in local minima, particularly when imaging non-sparse, dynamic objects through time-varying scattering that produces very low signal-to-noise ratio (SNR) speckles. Although subsequent improvements have been proposed to enhance convergence speed and imaging quality by incorporating additional spatial constraints or prior information \cite{hofer2018wide, lu2022single, li2022lensless, tahir2022adaptive}, they still face inherent limitations in reconstruction fidelity and computational efficiency.

Dynamic imaging through a static scattering medium within the memory effect range can be accomplished via an efficient one-step deconvolution process employing a pre-calibrated point spread function (PSF), akin to the method utilized in DiffuserCam. (see Fig.~\ref{fig:principle-illustrates}(b))~\cite{antipa2017diffusercam, chien2024space}. In this case, objects hidden behind the scattering medium can be simply reconstructed with the known PSF, which can be seen as solving a well-posed problem. However, the PSF measurement is relatively invasive due to the need to position a guidestar on the object side, making it impractical to image through time-varying scattering media. Numerous methods have been suggested for calibrating the system's transfer function in the absence of a guidestar in the past decade~\cite{wu2020non, yuan2020dynamic, zhang2023noninvasive, weinberg2024noninvasive, tu2025deep}. Most recently, I-CLASS has been proposed as a powerful method for imaging through dynamic unknown scattering~\cite{sunray2025matrix}, but it is based on the assumption of static objects. Similarly, recent image-guided computational holography methods also eliminate the need for a guidestar by optimizing a “virtual spatial light modulator (SLM)”. However, this computational optimization relies on a series of holographic (complex-field) measurements to digitally correct the aberration\cite{haim2025image}. Moreover, Multi-frame OTF Retrieval Engine (MORE)~\cite{yuan2020dynamic,lei2024dynamic} is a guidestar-free method for dynamic imaging through static scattering medium, which can actively calibrate the phase of OTF with a few shots of unknown dynamic object, instead of hundreds of frames in I-CLASS. Nevertheless, it has not been proven to perform effectively under dynamic scattering due to instability introduced by de-correlation of PSF in the time-varying imaging system. This type of "noisy" signal presents challenges for recovery using MORE, which relies on conventional nonlinear optimization techniques, despite its demonstration for low signal-to-noise ratio imaging in the context of imaging through scattering media.\cite{sun2023non,lei2024dynamic}. Recently, model-driven (or "untrained") neural networks have been employed to reduce the tendency of converging to local minima and demonstrate excellent noise resistance in iterative methods~\cite{wang2022far, tang2023deepsci, wang2024neuph, zhu2023displacement} by exploiting nonlinear optimization techniques and image prior. 

Motivated by these recent advances, we introduce NeOTF, a neural representation-based method for robustly retrieve the OTF of a scattering medium with a neural representation leveraging spatio-temporal memory effect. The NeOTF-enabled imaging pipeline is illustrated in Fig.~\ref{fig:principle-illustrates}(c) and Fig. ~\ref{fig:nn-otf-retrieval}. Our method is built upon the concept of implicit neural representation (INR) of the transfer function in linear imaging systems~\cite{sitzmann2020implicit, zhou2023fourier}. By leveraging the expressive power of INRs, NeOTF learns a spatially continuous model of the system’s OTF with only a few speckle images from unknown objects. We first validate NeOTF across a range of illumination bandwidths (up to a 300 nm broadband source), demonstrating state-of-the-art performance in both low-SNR and broadband dynamic imaging. Furthermore, NeOTF exhibits exceptional robustness, making it particularly effective for dynamic imaging scenarios in which object motion exceeds the evolution speed of the scattering medium (i.e., within the temporal memory effect range). In such cases, multiple speckle frames enable NeOTF to recover an “averaged” OTF for each time interval within the spatio-temporal memory effect range, ensuring consistent image reconstruction. This unique capability allows NeOTF to simultaneously manage time-varying scattering and dynamic samples, leveraging the spatio-temporal memory effect to continuously estimate the time-varying OTF and achieve high-fidelity video reconstruction under highly challenging scattering conditions.

\section{Methods}
\subsection{Physics model and constraints}

In the imaging through scattering setting, if the object is within the memory-effect range, the measured speckle intensity is well approximated by a convolution of the intensity of the object with the point spread function (PSF) of the imaging system, as the aberration is isoplanatic in the field of view~\cite{katz2014non},
\begin{equation}
    I \approx O\ast S,
    \label{eq:convolution}
\end{equation}
where $I$ and $O$ are the intensities of speckle and object, respectively. $S$ is the PSF of the imaging system. $\ast$ denotes the convolution operation. Based on this model, one of the most straightforward and commonly used methods for image reconstruction is deconvolution.

According to Eq.(\ref{eq:convolution}), the Fourier transform of the object can be expressed by the following equation:
\begin{equation}
\mathscr{F}\left( O \right) \approx  \frac{\mathscr{F}\left( I \right)}{\mathscr{F}\left( S \right)} = \frac{\left| \mathscr{F}\left( I \right) \right|}{\left| \mathscr{F}\left( S \right) \right|} e^{i\left( \phi_I - \phi_S \right)},
    \label{eq:speckle-inverse-filter}
\end{equation}
where $\mathscr{F}(\cdot)$ denotes the Fourier transform function, $\phi_I$ and  $\phi_S$ are the Fourier phases of $I$ and $S$, respectively. In an optical imaging system, $|\mathscr{F}(S)|$ essentially acts as a spatial low-pass filter. It has been reported that with only phase term $\phi_{S}$, the object can be recovered without the need of $|\mathscr{F}(S)|$\cite{yuan2020dynamic,mukherjee20183d}. So the Eq.(\ref{eq:speckle-inverse-filter}) can be further formulated as,
\begin{equation} 
\mathscr{F}\left( O \right) \approx \mathscr{F}\left( O^\prime \right) = \left| \mathscr{F}\left( I \right) \right| e^{i\left( \phi_I - \phi_S \right)},
\label{eq:phase-inverse-filter}
\end{equation}
where $O^\prime$ represents the low-pass filtered version of $O$ due to the diffraction limit of the imaging system. Therefore, subject to appropriate constraints, the object $O$ can be reconstructed as,
 \begin{equation} 
    O \approx O^{\prime} = \mathrm{RN}\left( \mathscr{F}^{-1}\left( \mathscr{F}\left( O^{\prime} \right) \right) \right)=\mathrm{RN}\left( \mathscr{F}^{-1}\left( \left| \mathscr{F}\left( I \right) \right| e^{i\left( \phi_I - \phi_S \right)}\right)\right),
    \label{eq:object-constraint}
\end{equation}
where $\mathscr{F}^{-1}(\cdot)$ denotes inverse Fourier transform function. $\mathrm{RN}(\cdot)$ represents the object domain constraints, which include the constraints of \emph{Support Region} and \emph{Real and Non-negative}~\cite{fienup1982phase}:
\begin{equation} 
    \mathrm{RN}\left(I(\mathbf{\rho})\right) =
\begin{cases}
  \Re(I(\mathbf{\rho})) & \text{if } I(\mathbf{\rho}) \ge 0 \text{ and } \mathbf{\rho} \in \mathcal{C} \\
  0 & \text{otherwise},
\end{cases}
\end{equation}
where $I(x,y)$ corresponds to the intensity on spatial position $\rho$, $\Re$ denotes the real part of complex value and $\mathcal{C}$ represents the position cluster of support region. This spatial constraint applied to reduce the solution space by restricting image reconstruction to a specific region.

To retrieve the phase terms, conventional phase retrieval algorithms, such as Error-Reduction (ER) and Hybrid-Input-Output (HIO), often struggle to recover accurate phase from noisy signals, especially under single-shot speckle constraints and simple error reduction algorithms, which are prone to local minima. In contrast, multi-frame speckles provide redundant constraints, which greatly reduce the solution space, leading to faster convergence and more reliable results. However, these traditional approaches still rely on iterative heuristics and often fail to generalize well under low signal-to-noise ratios or dynamic scattering conditions. Motivated by these limitations, we develop a more robust and learning-based framework for phase retrieval. Our proposed NeOTF leverages a gradient-descent-based optimization with a powerful optimizer that drives the solution towards higher accuracy even under noisy signal. The objective of the NeOTF training is to minimize the $L_1$ loss between the Fourier amplitudes of the measured speckles and the reconstructed images. The loss function, $\mathcal{L}$, is the sum of the $L_1$ losses of all speckles and the corresponding reconstruction results, expressed as:
\begin{equation}
    \mathcal{L}_{i}=\sum_j L_{1}\left(|\mathscr{F}(I_j)|,|\mathscr{F}(O_{\{i,j\}} ^\prime)|\right),
    \label{eq:loss}
\end{equation}
 where $i$ represents the $i$-th iteration during training and $j$ corresponds to the $j$-th speckle in the multi-frame speckles. The updated parameters $\theta^\prime$ are optimized by Adam optimizer\cite{2015-kingma}, 
\begin{equation}
\theta^\prime=\text{Adam}(\theta,\mathcal{L}_i),
    \label{eq:adam}
\end{equation}
where $\theta$ denotes the trainable parameters of our neural network (details are discussed in Sect. \ref{sec}).
\subsection{NeOTF-enabled imaging through scattering pipeline.}\label{sec}
\begin{figure}[htb]
    \centering
    \includegraphics[width=1.0\linewidth]{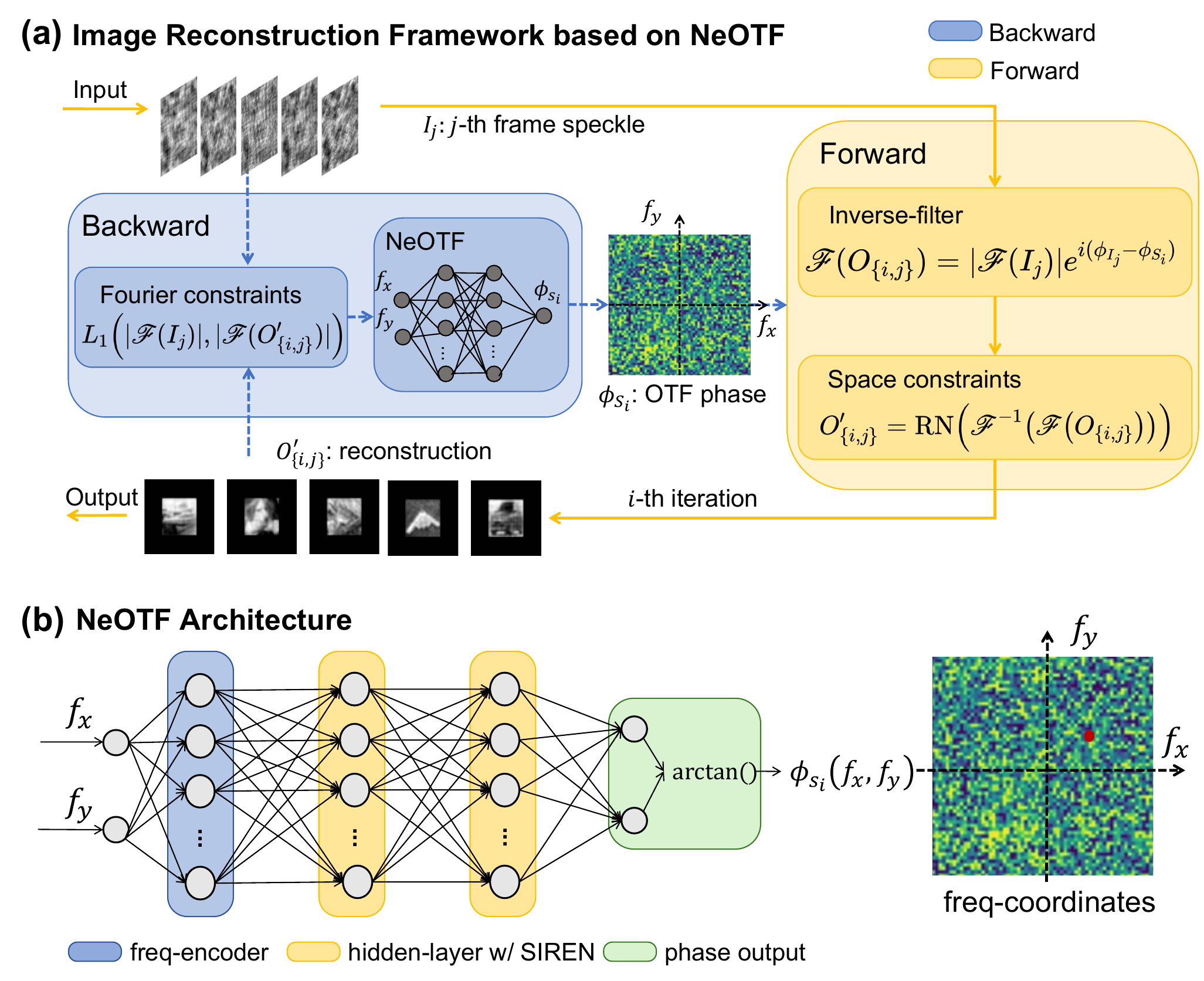}
    \caption{Overiew of our proposed NeOTF-enabled imaging framework. (a) Pipeline of NeOTF optimization and image reconstruction. The optimization begins with multi-frame images as input, and the reconstructed object intensity image(s) as the output. The yellow solid line indicates the forward pass in the image reconstruction workflow, while the blue dotted line depicts backward pass mainly composed of the OTF retrieval. (b) The network architecture of NeOTF. It consists of a frequency encoding layer, two hidden layers with SIREN activation functions, and an output layer with arctan activation function. The input is coordinates of $(f_x,f_y)$, and the output $\phi_{S_{i}}(f_x,f_y)$ is the calculated phase corresponding to coordinates $(f_x,f_y)$ at the $i$-th epoch.}
    \label{fig:nn-otf-retrieval}
\end{figure}

The imaging through scattering pipeline based on NeOTF is illustrated in Fig.~\ref{fig:nn-otf-retrieval}(a). The reconstruction pipeline operates in a self-supervised manner. In the training stage, a small set of speckle images is fed into the forward model, which outputs the predicted reconstructions. This prediction results are then compared with the input data in the Fourier domain to compute the loss. The proposed OTF retrieval process is an iterative optimization procedure designed to train the NeOTF network serving as a spatially continuous representation of the OTF's phase, $\phi_S$. Within $i$-{th} epoch, the algorithm processes a set of captured speckle frames $I$ to update its estimate of the OTF phase, $\phi_{S_i}$. For each individual speckle frame $I_j$, a forward physical model is applied that recovers objects through inverse filtering, denoted as $\mathrm{Forward}(\cdot)$. This function takes the Fourier transform of the speckle, $\mathscr{F}(I_j)$, and the current network-generated OTF phase, $\phi_{S_i}$, to reconstruct an estimate of the object, $O^{\prime}_{\{i,j\}}$. Next, a loss function is computed to quantify the consistency between the reconstruction and the physical measurement in the Fourier domain. We define this loss as the $L_1$ distance between the Fourier amplitude of the estimated object, $\left|\mathscr{F}(O^\prime_{\{i,j\}})\right|$, and that of the original speckle, $\left|\mathscr{F}(I_j)\right|$. The model iterates through a batch of frames, and the individual loss of each frame is calculated and then summed to form the accumulated losses, $\mathcal{L}_i$. This aggregate loss signal drives the update of the network's parameters via backpropagation, using the Adam optimizer. This entire process is repeated for a set number of epochs or until a convergence criterion is met, resulting in an optimized model of the system's OTF phase. In the inference stage, the fully trained model is used to make a final prediction of $\phi_{s_i}(f_x,f_y)$ with the frequency coordinates, $(f_x,f_y)$, as the input. The use of multiple images in NeOTF training is conceptually similar to the phase diversity method\cite{johnson2024phase,paxman1988optical}, where the undistorted object and the system's OTF are jointly estimated by measuring the distorted images of different objects through a common scattering system. The Fourier spectrum diversity of speckles not only enhances the completeness of OTF sampling, but also further constrains the solution space.

 As depicted in ~\ref{fig:nn-otf-retrieval}(b), the architecture of the NeOTF is based on a custom implicit neural representation (INR)\cite{sitzmann2020implicit} that encodes data as continuous functions parameterized by a neural network, rather than discrete samples. It is essentially a simple multilayer perceptron with a frequency encoder layer, followed by two hidden layers each containing 128 units, and then an activation function module based on SIREN activation functions~\cite{tancik2020fourier, sitzmann2020implicit} and phase output layer that maps the input of Fourier-domain frequency coordinates to the output OTF phase that ranges between $(-\pi, \pi)$ continuously. Compared with network backbones based on the Deep Image Prior (DIP)~\cite{ulyanov2018deep, zhou2020diffraction}, which are based on convolutional architectures and considers the input as a discrete representation of a signal with a fixed size, INR is more flexible and can be used to handle the inputs of different image sizes without additional adjustment to the architecture. Similar to common INRs in the two-dimensional domain, the network takes coordinates in the two-dimensional frequency domain as input and outputs the corresponding phase value. It is worth noting that the phase of the OTF in a strong scattering imaging system usually exhibits a statistical distribution similar to high-frequency noise. To help retrieve the such kind of phases, we apply a additional positional encoder layer to enhance INR's ability to fit high-frequency signals. The phase map generated by the optimized network's OTF output is directly applicable for reconstructing the object from the speckle image.

The proposed method is effective not only for time-invariant OTFs when imaging through a static scattering medium, but also for reconstructing previously unseen dynamic objects through dynamic scattering, an often overlooked challenge that is rarely addressed in prior work yet frequently encountered in real-world applications. The latter can be achieved through in situ ‘self-recalibration’ or re-optimization using continuously acquired data, without the need to reintroduce guidestars for intermediate calibration or characterization of the scattering medium.

\section{Results}
\label{sect:results}
\subsection{Dynamic imaging through scattering under low SNR and broadband illumination scenario}

\begin{figure}[!htb]
    \centering
    \includegraphics[width=1.0\linewidth]{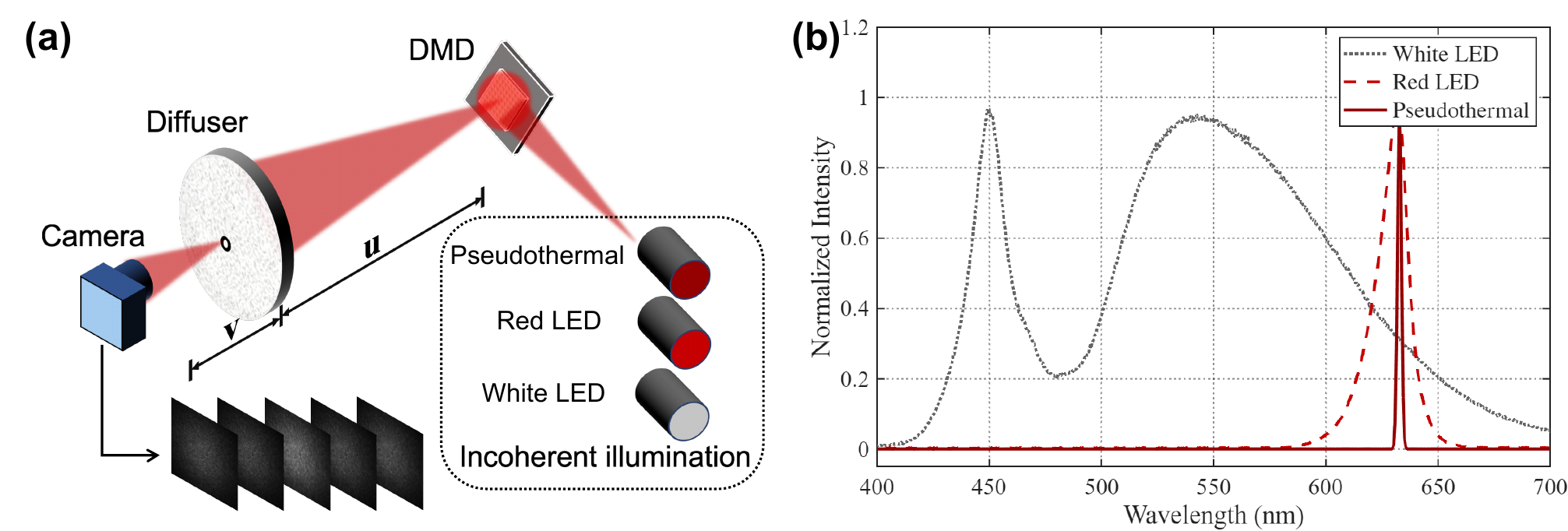}
    \caption{Experimental setup and illumination source characterization (a) Experimental setup of non-invasive imaging through a scattering medium. In our experimental demonstration, we use three different incoherent illumination sources, the first one is a rotating ground glass (not shown in the figure) combined with a He-Ne laser, which is used to generate a pseudothermal light source; the second one is a red monochromatic LED; and the third one is a white LED. A  Digital Micromirror Device (DMD) is used to display the dynamic object, which reflects the light through a diffuser and create speckle image captured by a camera. (b) Spectral intensity distribution of the sources in (a) measured by an optical fiber spectrometer.}
    \label{fig:experimental-setup}
\end{figure}

By leveraging multi-frame speckle constraints, NeOTF substantially narrows the solution space for faster convergence compared to traditional phase retrieval algorithms that rely on a single speckle frame. This constrained optimization is hypothesized to improve the robustness of NeOTF's reconstructions in low SNR scenarios. To validate this, we conducted dynamic imaging through scattering experiments under broadband illumination, a common low-SNR condition where speckles exhibit low contrast, blurred features, and strong background noise.

\begin{figure}[!htb]
    \centering
    \includegraphics[width=1.0\linewidth]{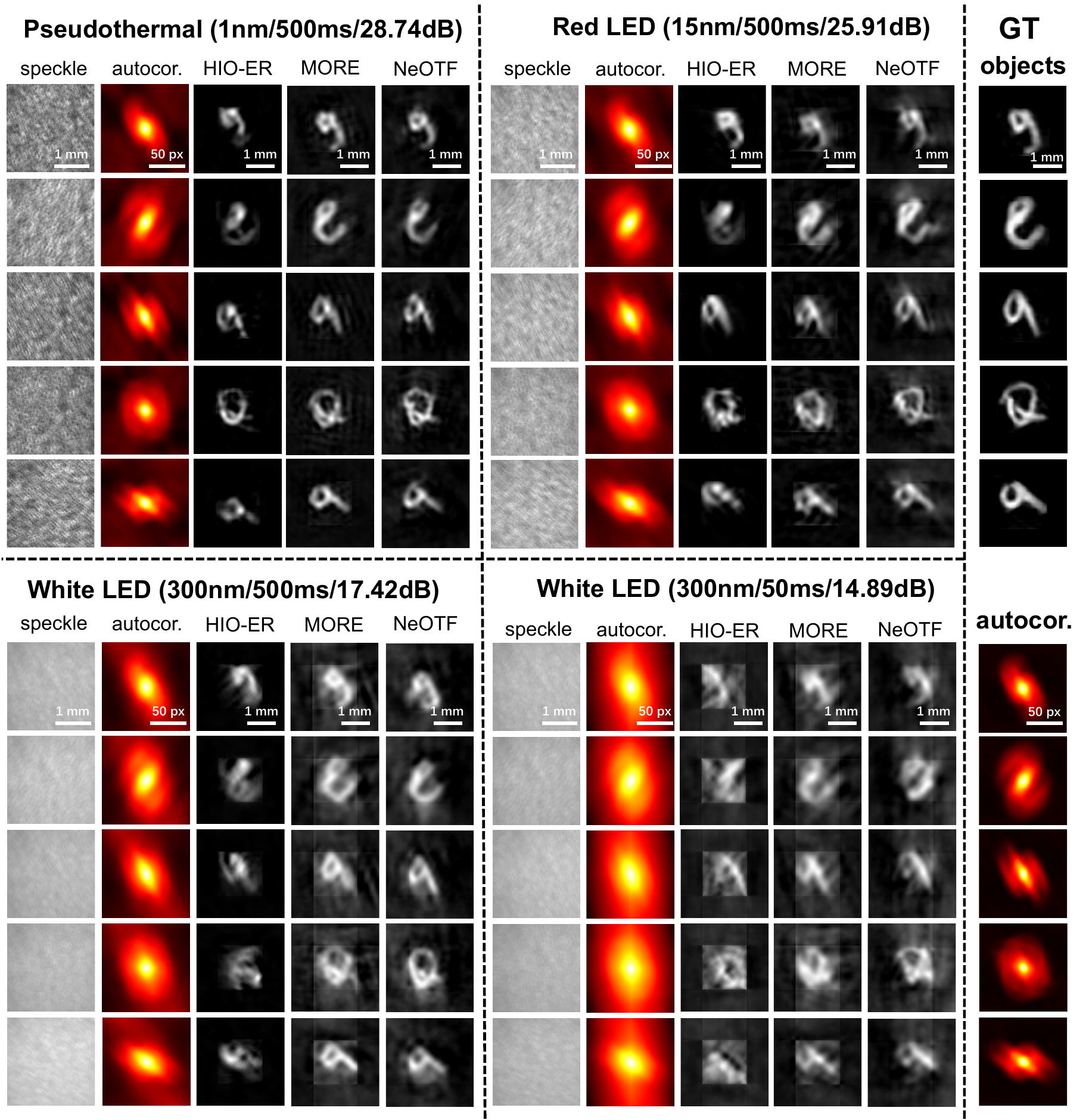}
    \caption{Experimental validation of NeOTF's robustness in low-SNR and incoherent illumination conditions. Reconstructions are compared under several different incoherent illumination sources and exposure times: a He-Ne laser with rotating diffuser as a pseudothermal light source (1~nm / 500~ms), a red LED (15~nm / 500~ms), and a white-light LED (300~nm / 500~ms and 50~ms), yielding the autocorrelation's SNRs of 28.74~dB, 25.91~dB, 17.42~dB, and 14.89~dB, respectively. Each panel shows the raw speckle image measured through scattering, its 2D autocorrelation image, the reconstruction results obtained based on conventional methods including the traditional HIO-ER and MORE, and the result from our proposed NeOTF method, with the ground truth (GT) image on the right. autocor.: 2D autocorrelation of the speckle image.}
    \label{fig:lowSNR-result}
\end{figure}

Our experimental setup for imaging with low SNR speckles is schematically illustrated in Fig.~\ref{fig:experimental-setup}(a). A He-Ne laser (central wavelength $\lambda = 632.8$~nm, FWHM=1~nm) combined with a rotating ground glass to generate a pseudothermal light source, a red monochromatic LED (FWHM=15~nm) and a white LED (FWHM=300~nm) are served as the illumination source (see Fig.~\ref{fig:experimental-setup}(b)), respectively. A Digital Micromirror Device (DMD; F4100, 0.7" XGA), with a resolution of $1024 \times 768$~pixels and a pixel pitch of $13.68$~$\mu$m, was used to display dynamic patterns as the 2D objects. The DMD was positioned at a distance of $u = 635$~mm from the scattering medium. The scattering medium consisted of a 220-grit ground glass diffuser with an 8~mm diameter aperture placed directly in front of it allowing us to perform lensless imaging. A monochrome CMOS camera (Daheng Imaging MER-2000-19U3M; $5496 \times 3672$~pixels, $2.4$~$\mu$m pixel pitch size), positioned at a distance of $v = 100$~mm behind the diffuser, was used to capture the resulting speckle patterns with different illumination sources.

The raw captured speckle images then underwent a digital pre-processing pipeline. First, a combination of frequency and Gaussian filtering was applied to reduce noise. Subsequently, a central region of $1024 \times 1024$~pixels was cropped from each filtered image and down-sampled to a final resolution of $512 \times 512$~pixels for subsequent denoising and filtering. The corresponding speckle patterns captured for each source and its autocorrelation after processing are shown in Fig.~\ref{fig:lowSNR-result} and those images have been normalized for visualization purposes. We use speckle autocorrelation image as a proxy for characterizing speckle image's SNR, as it is directly linked to the object’s spatial-frequency content. A higher similarity between the speckle's autocorrelation and the object's true autocorrelation indicates that more object-relevant information is contained in speckles, which can be interpreted as a higher information-effective SNR. As the spectral bandwidth of the laser source increases, the speckle autocorrelation becomes over-smooth, and the spatial information of the object is then hidden in the low-frequency noise, which are more likely to mislead the error reduction direction of the optimization algorithms. Imaging under these conditions has long been a central challenge in broadband dynamic imaging through scattering media. To assess performance, we benchmark our reconstructions against most widely used algorithm: HIO-ER (Hybrid-Input-Output \& Error-Reduction) and one of the state-of-the-art low-SNR imaging methods: MORE, while systematically varying exposure time and laser spectral bandwidth to produce a controlled range of SNRs.

In Fig.~\ref{fig:lowSNR-result}, results under narrow-band illumination, both the NeOTF and HIO-ER algorithms can accurately reconstruct the object's structure with high clarity. However, under broadband illumination, the contrast of both the speckle patterns and their corresponding autocorrelations is significantly reduced, causing the object's spatial information to be overwhelmed by low-frequency background noise. This poses a significant challenge for traditional phase retrieval algorithms, which struggle to recover a credible image from such overly smooth signals. For these algorithms, the solution space is vast, making efficient reconstruction from a single, weakly-constrained speckle image difficult. In contrast, NeOTF, a neural representation of the system's transfer function, is inherently designed to leverage constraints from multiple speckle frames. This multi-frame approach effectively narrows the possible solution space, thereby enhancing both the efficiency and quality of the final reconstruction. Notably, the reconstruction quality of NeOTF remains remarkably stable across all tested conditions, demonstrating a strong resilience to the SNR degradation caused by increasing spectral bandwidths. MORE has been one of the state-of-the-art methods for imaging under low SNR scenario and NeOTF inherits the powerful constraints of multi-frame speckle images that also underlie the effectiveness of MORE. Additionally, benefiting from neural network–based optimization, NeOTF substantially outperforms traditional nonlinear optimization methods such as MORE, particularly in handling background noise under extremely low-SNR conditions down to 14.89~dB (see panels "White LED (300~nm\&500~ms)" and "White LED (300~nm \& 50~ms)" in Fig.~\ref{fig:experimental-result}).

\subsection{Dynamic imaging of C. elegans embryogenesis and chiral rotation through static scattering medium}

Dynamic imaging through scattering media can be categorized into two types: time-varying scattering, where the medium changes and time-invariant scattering, where the medium remains almost fixed but the object itself changes over time. Imaging dynamic objects through dynamic scattering media is a challenging task due to the decorrelation and uncertainty of speckle caused by the spatio-temporal variations from simultaneous motion of both the object and the scattering medium.

\begin{figure*}[!htb]
    \centering
    \includegraphics[width=0.6\linewidth]{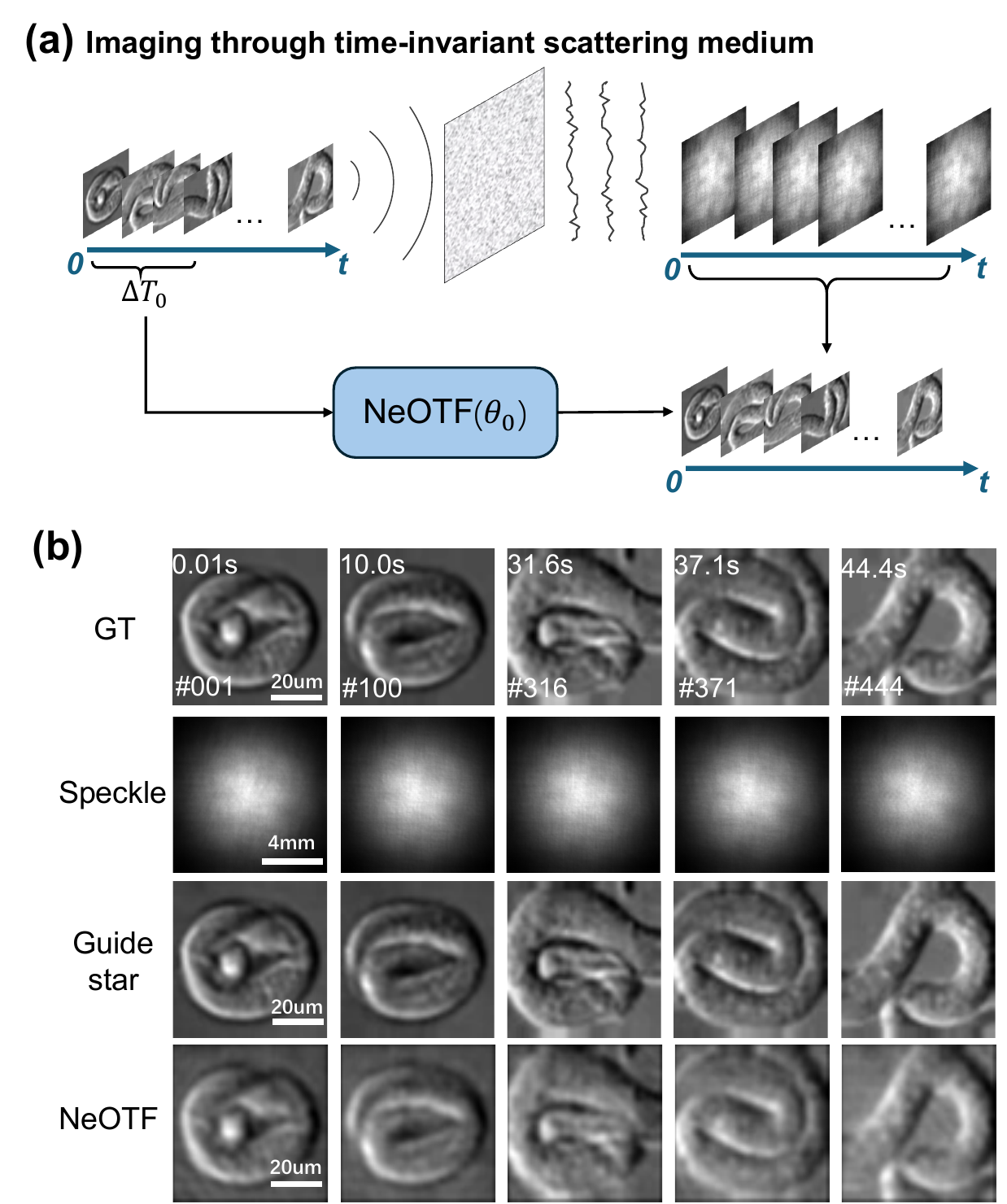}
    \caption{Dynamic imaging of a \textit{C. elegans} embryogenesis and chiral rotation through static scattering medium. (a) Illustration for dynamic imaging through a static diffuser. The OTF of the time-invariant imaging system is retrieved from initial 5 frames of speckles in a short time period ($\Delta T_0$). Subsequent dynamic scenes are all reconstructed using the restored NeOTF ($\theta_0$). (b) Reconstructed results from the dynamic speckles from simulated guidestar-based reconstruction and the proposed NeOTF method. "GT" indicates ground truth images. "Speckle" indicates low SNR speckle images collected through scattering. "guidestar" indicates the results recovered from a simulated OTF measurement using a synthetic guidestar. "NeOTF" shows the images reconstructed by the proposed guidestar-free based NeOTF framework.
    }
    \label{fig:dynamic_imaging_simulation}
\end{figure*}

Because NeOTF can recover the OTF of an imaging system in situ, even for scattering imaging systems with time-invariant OTFs, each frame can be reconstructed using a single inverse filtering step. To validate NeOTF's performance on dynamic imaging through static medium, we conducted a dynamic imaging simulation using a 45 seconds microscopic video of a C. elegans embryonic chiral rotation\cite{Hall2017-sx} during its late development, including 450 frames of dynamic scenes. Studying C. elegans chiral rotation offers insight into the physical principles of symmetry breaking and developmental organization in living systems. Such rapid chiral rotation happens as body-wall muscles become active, with motion occurring at tens of frames per second (see Fig.~\ref{fig:dynamic_imaging_simulation}(a)). Each video frame was down-sampled to $60\times60$~pixels to serve as ground truth. These images were then used to generate $512\times512$~pixel speckle patterns through a 20$\times$ magnification. We used a random selection of 5 speckle frames as input to train NeOTF. After 10,000 iterations, the results show that while NeOTF accurately restores the overall shape of the C. elegans embryo and main internal structures, there is a loss of high-frequency details compared to the ground truth and the diffraction-limited images (see the Movie~1 in supplementary material). This is an expected outcome, as the neural network's optimization process prioritizes fitting the low-frequency components that dominate the loss function, often at the expense of finer, high-frequency signals.

\subsection{Dynamic imaging of C. elegans embryogenesis and chiral rotation through dynamic scattering medium}

\begin{figure*}[!htb]
    \centering
    \includegraphics[width=1.0\linewidth]{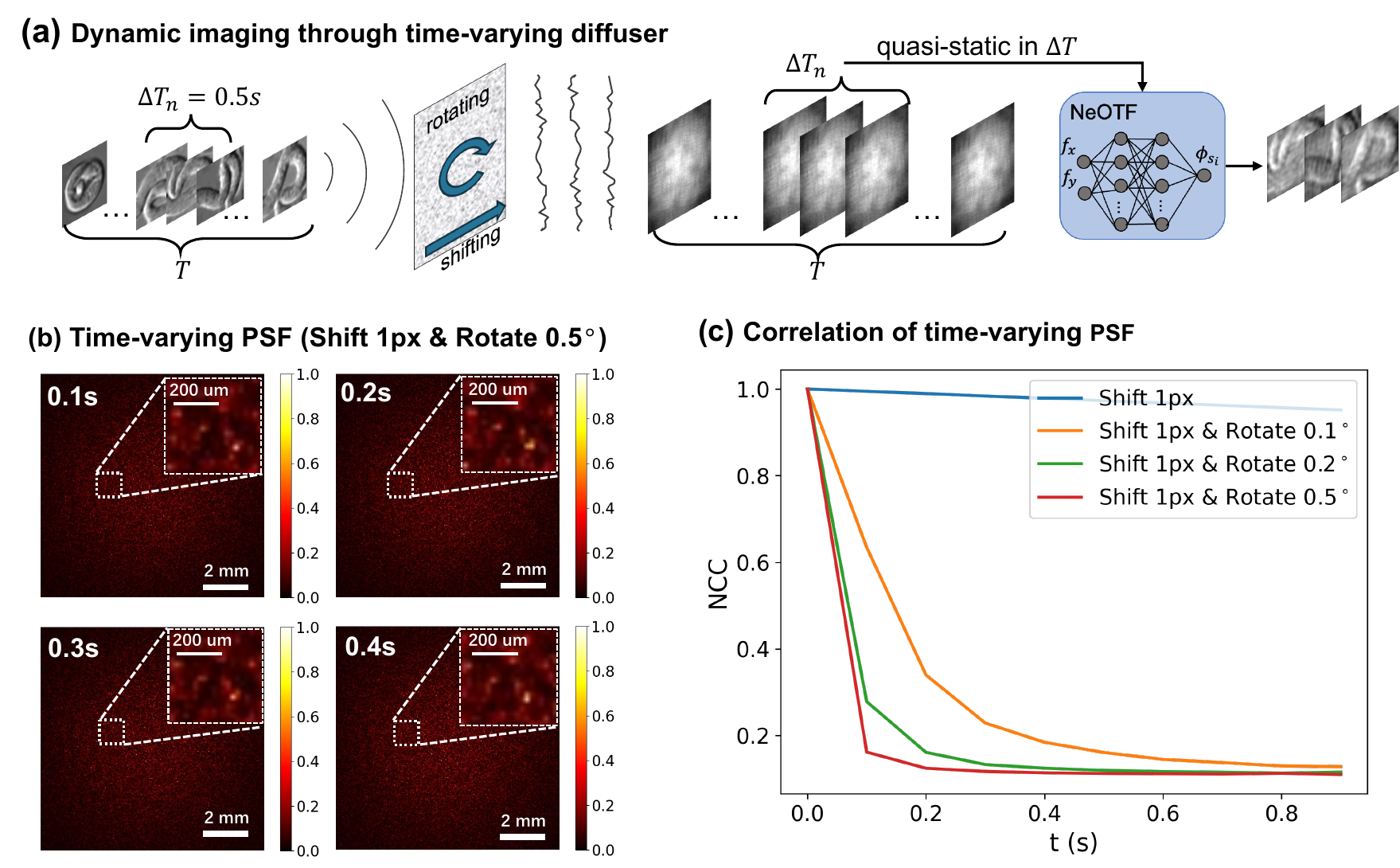}
    \caption{Dynamic imaging through dynamic scattering. (a) NeOTF-enabled dynamic imaging through dynamic scattering medium within temporal memory effect range. (b) Characterization of time-varying speckled PSF at different time. The speckles in the white dashed boxed are zoomed in for visualization. (c) Characterization of temporal memory effect by analyzing correlation decay of time-varying PSF. "NCC" represents the correlation between the PSF at time $t$ and the PSF at time $t=0$. The dynamics of the scattering medium are analyzed under different conditions: a lateral shift of 1 pixel (px) and rotations of $0^{\circ}$, $0.1^{\circ}$, $0.2^{\circ}$, $0.5^{\circ}$ per 0.1 s. As the rotation per frame increases, the PSF decorrelates more rapidly, resulting in a narrower temporal memory-effect range.
    }
    \label{fig:dynamic_scattering_experiments}
\end{figure*}

For time-varying scattering media within a short time window, it is worth noting that, although the scattering medium phase varies slightly from frame to frame, it remains highly temporally correlated, i.e. within the temporal memory-effect range (as characterized in Fig. \ref{fig:dynamic_scattering_experiments}c). This assumption enables NeOTF to reconstruct reliable images within a short time window using temporally adjacent speckle patterns. Due to NeOTF's noise tolerance and robustness, it is capable of imaging in such dynamic scenarios (see Fig.\ref{fig:dynamic_scattering_experiments}(a)). To verify NeOTF's imaging capabilities under time-varying scattering, we spatially shifted and rotated the scattering medium phase in continuously (1~pixel horizontal translation and $0.5^{\circ}$ rotation per frame) to simulate the phase change during a short time period, as shown in Fig.~\ref{fig:dynamic_scattering_experiments}(b). 

\begin{figure*}[!htb]
    \centering
    \includegraphics[width=1.0\linewidth]{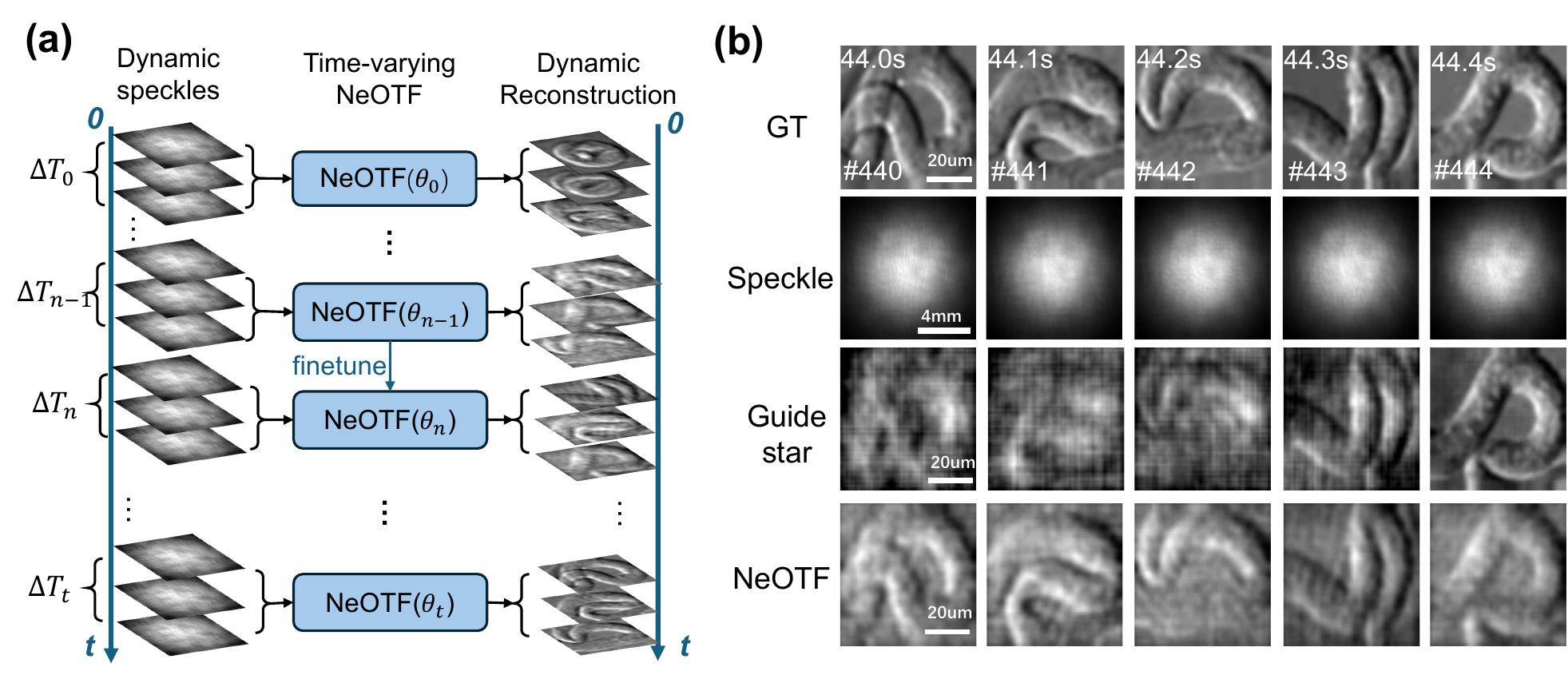}
    \caption{Dynamic imaging of C. elegans embryogenesis and chiral rotation through dynamic scattering medium. (a) Pipeline of NeOTF-enabled imaging through time-varying scattering medium. NeOTF($\theta_0$) is optimized from initial time period $\Delta T_0$ and subsequent weights $\theta_{n}$ is fine-tuned from the weights $\theta_{n-1}$ of last time checkpoint of $\Delta T_{n-1}$. (b) C. elegans embryogenesis and chiral rotation reconstructed from the dynamic speckles. "guidestar" indicates the results recovered from a single OTF measurement using a guidestar. "NeOTF" shows the images reconstructed by the proposed guidestar-free based NeOTF framework.
    }
    \label{fig:dynamic_scattering_results}
\end{figure*}

The pipeline of imaging through dynamic scattering medium is shown as Fig.\ref{fig:dynamic_scattering_results}(a). For the time-varying NeOTF, we first train the NeOTF($\theta_0$) for 5000 epochs as an initial step, where $\theta_0$ means the weights in the time period $\Delta T_0$. Then for the next short period of time (a few acquisitions of images), the weights of $\theta_0$ were loaded to finetune the NeOTF($\theta_1$) with the speckles of the next period of time. The model finetuning process requires much less epochs to reach the convergence due to the highly correlated OTF between two adjacent time points. The weights from the previous NeOTF iteration serve as a good starting point for model optimization at the next time step. The finetuning of the NeOTF at time $t=n$ can be expressed as
\begin{equation}
    \mathbf{NeOTF}(\theta_{n})=\mathbf{FT}\left(\theta_{n-1},I(\Delta T_n)\right),
\end{equation}
where $\theta_{n}$ are the weights of NeOTF for the time window $\Delta T_0$, $I(\Delta T_0)$ are the speckles during $\Delta T_0$ and $\mathbf{FT}$ means the fine-tuning process with weights $\theta_{n}$.

In our simulations, we train the initial start point for five thousands of epochs and fine-tune each subsequent time point for five hundreds of epochs, which only need 10\% of training from random initialization to reach the convergence. We divide the continuous speckle in the time domain into subsets of 5 frames, assuming that these 5 frames are within a pre-defined range of highly correlated PSF changes. As shown in Fig.~\ref{fig:dynamic_scattering_results}(b), the "NeOTF" is the result of reconstructing these five speckles using the same OTF retrieved by NeOTF (The video of reconstruction can be found in Movie~2 of the supplementary material). For comparison, the 'guidestar' image was generated by reconstructing the five speckle frames. The OTF used for this reconstruction was measured from the last speckle frame using a guidestar based method. Naturally, the best results appear between the current frame and the adjacent frames (e.g., 4 and 5th frame), and the quality of reconstruction diminishes as the correlation of the Point Spread Function (PSF) decreases, which is associated with a reduction of PSF correlation away from the range of the temporal memory effect (see Fig.\ref{fig:dynamic_scattering_experiments}(c)). Comparatively, since each iterative update of NeOTF is based on the reconstruction results of all frames, so the reconstruction quality of each frame is relatively stable in NeOTF. These results demonstrate that NeOTF can perform dynamic imaging through both time-varying and time-invariant scattering media.

\section{Discussion}
\label{sect:discussion}
In NeOTF, the INR framework effectively models the spatially continuous and complex characteristics of the optical transfer function (OTF), enabling high-fidelity image reconstruction beyond traditional approaches. With the retrieved OTF, objects can be reconstructed by simple inverse filtering. The imaging process of NeOTF has high temporal resolution and efficiency and does not require a guidestar or additional optical components like wavefront shapers. Besides that, as a universal optimization framework, other constraints can be incorporated, such as Total Variation and Sparse Representation, which would improve the quality and convergence of NeOTF.

For computational imaging systems, particularly edge devices deployed in real-world scenarios, computational efficiency and generalization are key metrics for practical application. In the following discussion, we first quantitatively analyze the computational efficiency of NeOTF and propose strategies to improve it without significantly compromising image quality, using limited sampling in the frequency domain. Second, we evaluate the generalization performance of NeOTF across scattering scenes with varying anisotropy factors, to understand its limitations under different scattering properties.

\subsection{Computational efficiency}

Training coordinate-based implicit neural representations is notoriously intensive \cite{mildenhall2021nerf, shen2021non}. This demand arises because the underlying multilayer perceptron (MLP) must process dense coordinates in the spatial domain during each training iteration. A key advantage of NeOTF lies in its computational efficiency, which stems from its novel operation in the frequency domain. This approach offers two distinct benefits:

First, because the imaging system's PSF is real-valued, its Fourier transform, the OTF, must exhibit Hermitian symmetry. Consequently, NeOTF only needs to sample half of the frequency coordinates to represent the entire phase map. This strategy inherently doubles the computational efficiency and further constrains the solution space, which improves convergence. Second, NeOTF capitalizes on the principle that natural images are inherently sparse in the frequency domain, with most significant information concentrated in the low-frequency components. In contrast, the high-frequency components of a speckle pattern are often dominated by signal-independent sensor noise. This allows us to train the network efficiently by sampling only a sparse subset of coordinates within the central low-frequency region of the phase map. This reduction in NeOTF's input dimensionality creates a direct trade-off between reconstruction quality and computational speed. As demonstrated by our experimental results, for simple objects such as numerical digits, sampling as little as 7.0~\% of the frequency coordinates in OTF is sufficient to maintain high-fidelity reconstruction while significantly accelerating the training process.

\begin{figure}[!htb]
    \centering
    \includegraphics[width=1.0\linewidth]{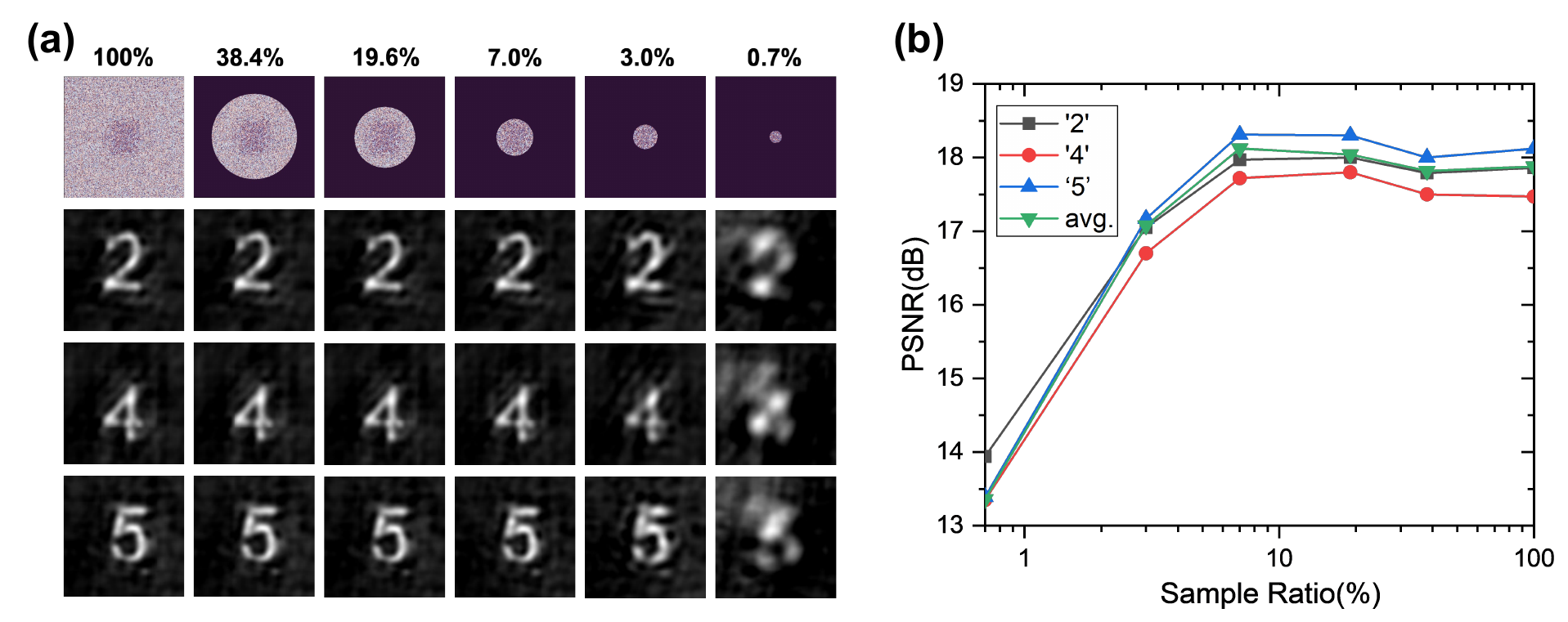}
    \caption{Experimental results of NeOTF reconstruction under different sampling ratio in the spatial frequency domain. The speckle images were captured in the experiments shown in Fig.\ref{fig:experimental-setup}(a). (a) Phase map and reconstructed images. The first row of images show the phase map of the system's OTF $\phi_s $ retrieved by NeOTF at different sampling ratios. Rows 2 to 4 are the object reconstruction results, respectively. (b) Quantification of PSNR (peak signal-to-noise ratio) of the reconstructed images of objects ('2', '4', '5') under various sampling ratios.}
    \label{fig:experimental-result}
\end{figure}

\begin{table}[htb]
\centering

\renewcommand{\arraystretch}{1.0}
\begin{tabular}{cccccc}
   \toprule
   Sampling Ratio (\%)& Input Size & Time (s) & Memory (GB) & GFLOPs & PSNR (dB) \\
   \midrule
   100 & (131584, 2) & 14.5 & 1.4 & 9.7 &  17.87\\
   38.4 & (50602, 2) & 9.3 & 0.8 & 3.77  & 17.81\\
   19.6 & (25900, 2) & 8.1 & 0.7 & 1.90  & 18.04\\
   7.0 & (9386, 2) &  7.8 & 0.6 & 0.67  & 18.12\\
   3.0 & (4206, 2) & 7.2 & 0.4 & 0.29 & 17.08\\
   0.7 & (1080, 2) & 7.1 & 0.4 & 0.07 & 13.36\\
   \bottomrule
\end{tabular}
\vspace{5pt}
\caption{Computational cost of training NeOTF under various sampling ratios in the spatial frequency domain. "Input Size" is the coordinate dimensions provided to NeOTF.  "Time" refers to the time spent on 5000 iterations of training. “Memory” denotes the GPU Video RAM (VRAM) usage, “GFLOPs” (Giga Floating-Point Operations per Second) is computed from the model’s inference. "PSNR" indicates the PSNR quantification of the reconstructed images.}
\label{table}
\end{table}

In our experiments, we only used an NVIDIA RTX 4090 laptop GPU to perform NeOTF reconstruction on speckle patterns with a resolution of 512$\times$512 pixels. We evaluated the computational resources and time consumed at various sampling ratios, with the detailed results presented in Table~\ref{table}. In practice, NeOTF implements its efficiency-enhancing sparse sampling strategy by confining its training to a circular region within the low-frequency spectrum. As demonstrated in Fig.~\ref{fig:experimental-result}, this approach enables a direct trade-off between reconstruction quality and speed. For simple objects such as the digits used in our experiments, reducing the sampling rate to just 7.0~\% yields a 100~\% increase in training speed with no discernible loss in image quality. This adaptability to balance fidelity with computational cost makes NeOTF a strong candidate for deployment on resource-constrained platforms, such as edge computing devices.

\subsection{Generalization to anisotropic scattering medium}

Because NeOTF uses only the OTF phase for image reconstruction, its generalization capability for anisotropic scattering media remains unexplored. Here, we simulated random phases on surfaces with varying anisotropic factor $\alpha$ of scattering and attempted to use NeOTF to reconstruct images through scattering media. Higher anisotropic factor $\alpha$ means more forward-directed scattering, resulting in speckle patterns that are less diffused and more spatially correlated.

\begin{figure}[!htb]
    \centering
    \includegraphics[width=1.0\linewidth]{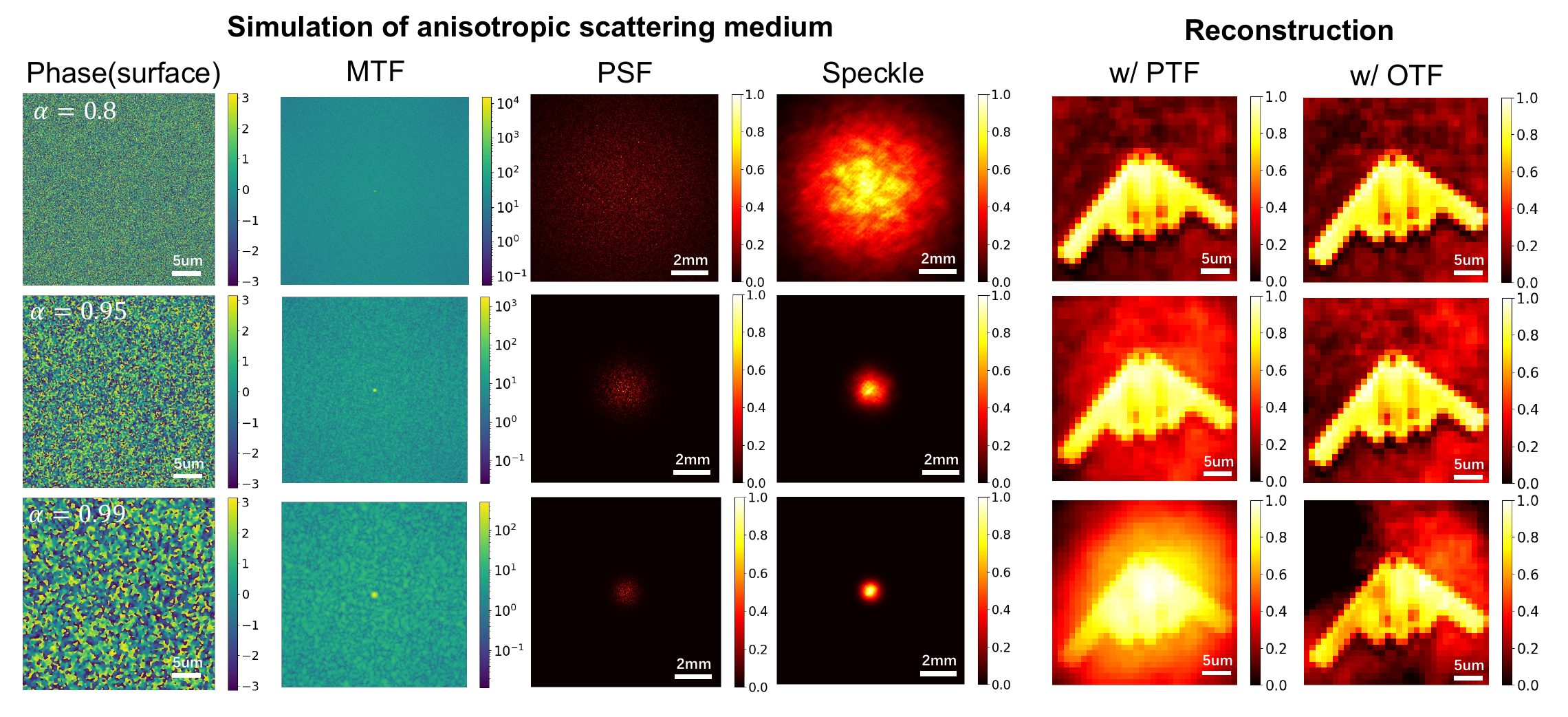}
    \caption{OTF inverse filtering imaging results for different random phase distributions. "Phase (surface)" represents the random phase distribution on the scattering surface under different anisotropic factor $\alpha$. MTF, PSF, and object speckle indicate the corresponding imaging system's modulation transfer function, point spread function, and the resulting speckle pattern, respectively. The last two columns show the reconstruction results using only the phase of OTF (PTF) and the full OTF (including both PTF and MTF).}
    \label{fig:roughness-diffusers}
\end{figure}

The final two columns of Fig.~\ref{fig:roughness-diffusers} show the objects reconstructed via direct inverse-filtering, using the phase of OTF (PTF) alone and the full OTF, respectively. Notably, the highest-fidelity reconstruction with the PTF is achieved in $\alpha = 0.8$. Conversely, the poorest result occurs in $\alpha=0.99$, the most weakly scattering medium in our simulations. This discrepancy arises because the physical model underlying NeOTF assumes a strongly scattering medium. Under this assumption, as shown in Eq.~\eqref{eq:speckle-inverse-filter}, the magnitude of the scattering spectrum, $|\mathscr{F}(S)|$ (MTF), is approximated as a low-pass filter with a uniform frequency response and can therefore be disregarded. In weakly scattering media, however, this non-uniform frequency response of MTF distorts the object's amplitude. Consequently, for weakly scattering scenarios, even the true OTF phase is insufficient to fully recover the object's information.

The current framework performs well with anisotropic factor of $\alpha = 0.95$ or below, which covers already a wide range of applications such as biological imaging where anisotropic factor is around $0.9$. Beyond $0.95$ generalization has limitations that suggest avenues for future research. The method is presently designed for strongly scattering media, and its adaptation to weakly scattering environments remains a further optimization. Moreover, the random-like distribution of the OTF phase in strong scattering systems limits the network’s ability to the high-frequency fitting capability. Exploring novel network architectures or more efficient phase encoding methods could further enhance NeOTF's scalability.

\section{Conclusion}

In conclusion, we have introduced NeOTF, a guidestar-free method that leverages an implicit neural representation to retrieve the system's OTF. This approach is designed to overcome the challenge of ill-posedness inherent in traditional phase retrieval methods and speckle decorrelation in dynamic imaging. Validated by both numerical simulations and physical experiments, our work demonstrates that NeOTF significantly improves reconstruction fidelity by enforcing robust physical constraints leveraging spatio-temporal memory effect from multi-frame speckles. By operating without guidestars or other invasive priors, and by capitalizing on a simple and efficient central frequency-domain sampling, NeOTF achieves both robust and computationally efficient image reconstruction, establishing it as a powerful tool for dynamic imaging through both time-invariant and time-varying media even under broadband, lensless, and low-SNR condition.

\subsection* {Acknowledgments}
We thank H. Shroff for inspiring us to look into the C. elegans developmental dynamics. F. X. acknowledges support from startup funds from the Samueli School of Engineering at the University of California, Irvine.


\bibliography{sample}   
\bibliographystyle{spiejour}   





\end{spacing}
\end{document}